\documentclass[english]{article}
\usepackage[T1]{fontenc}
\usepackage[latin9]{inputenc}
\usepackage{amssymb}
\usepackage{graphicx}
\usepackage{esint}
\usepackage{babel}
\begin{document}

\title{\textbf{Quantum Hair on Colliding Black Holes}}

\author{\textbf{Lawrence Crowell$^{1,+}$ and Christian Corda $^{2,*}$ }}

\maketitle
$^{1}$AIAS, Budapest, Hu/Albuquerque, NM/ Jefferson, TX 75657

$^{2}$Department of Physics, Faculty of Science, Istanbul University,
Istanbul 34134, Turkey and International Institute for Applicable
Mathematics and Information Sciences, B.M., Birla Science Centre,
Adarshnagar, Hyderabad 500063, India

E-mails: $^{+}$\emph{goldenfieldquaternions@gmail.com}; $^{*}$\emph{cordac.galilei@gmail.com}\}
\begin{abstract}
Black hole collision produce gravitational radiation which is generally
thought in a quantum limit to be gravitons. The stretched horizon
of a black hole contains quantum information, or a form of quantum
hair, which in a coalescence of black holes participates in the generation
of gravitons. This may be facilitated with a Bohr-like approach to
black hole (BH) quantum physics with quasi-normal mode (QNM) approach
to BH quantum mechanics. Quantum gravity and quantum hair on event
horizons is excited to higher energy in BH coalescence. The near horizon
condition for two BHs right before collision is a deformed $AdS$
spacetime. These excited states of BH quantum hair then relax with
the production of gravitons. This is then argued to define RT entropy
given by quantum hair on the horizons. These qubits of information
from a BH coalescence should then appear in gravitational wave (GW)
data. This is a form of the standard $AdS/CFT$ correspondence and
the Ryu-Takayanagi (RT) formula. 
\end{abstract}

\section{Introduction}

Quantum gravitation suffers primarily from an experimental problem.
It is common to read critiques that it has gone off into mathematical
fantasies, but the real problem is the scale at which such putative
physics holds. It is not hard to see that an accelerator with current
technology would be a ring encompassing the Milky Way galaxy. Even
if we were to use laser physics to accelerate particles the energy
of the fields proportional to the frequency could potentially reduce
this by a factor of about $10^{6}$ so a Planck mass accelerator would
be far smaller; it would encompass the solar system including the
Oort cloud out to at least $1$ light years. It is also easy to see
that a proton-proton collision that produces a quantum BH of a few
Planck masses would decay into around a mole of daughter particles.
The detection and track finding work would be daunting. Such experiments
are from a practical perspective nearly impossible. This is independent
of whether one is working with string theory or loop variables and
related models.

It is then best to let nature do the heavy lifting for us. Gravitation
is a field with a coupling that scales with the square of mass-energy.
Gravitation is only a strong field when lots of mass-energy is concentrated
in a small region, such as a BH. The area of the horizon is a measure
of maximum entropy any quantity of mass-energy may possess \cite{key-2},
and the change in horizon area with lower and upper bounds in BH thermodynamic
a range for gravitational wave production. Gravitational waves produced
in BH coalescence contains information concerning the BHs configuration,
which is argued here to include quantum hair on the horizons. Quantum
hair means the state of a black hole from a single microstate in no-hair
theorems. Strominger and Vafa \cite{key-3} advanced the existence
of quantum hair using theory of D-branes and STU string duality. This
information appears as gravitational memory, which is found when test
masses are not restored to their initial configuration \cite{key-4}.
This information may be used to find data on quantum gravitation.

There are three main systems in physics, quantum mechanics (QM), statistical
mechanics and general relativity (GR) along with gauge theory. These
three systems connect with each other in certain ways. There is quantum
statistical mechanics in the theory of phase transitions, BH thermodynamics
connects GR with statistical mechanics, and Hawking-Unruh radiation
connects QM to GR as well. These are connections but are incomplete
and there has yet to be any general unification or reduction of degrees
of freedom. Unification of QM with GR appeared to work well with holography,
but now faces an obstruction called the firewall \cite{key-5}.

Hawking proposed that black holes may lose mass through quantum tunneling
\cite{key-6}. Hawking radiation is often thought of as positive and
negative energy entangled states where positive energy escapes and
negative energy enters the BH. The state which enters the BH effectively
removes mass from the same BH and increases the entanglement entropy
of the BH through its entanglement with the escaping state. This continues
but this entanglement entropy is limited by the Bekenstein bound.
In addition, later emitted bosons are entangled with both the black
hole and previously emitted bosons. This means a bipartite entanglement
is transformed into a tripartite entangled state. This is not a unitary
process. This will occur once the BH is at about half its mass at
the Page time \cite{key-7}, and it appears the unitary principle
(UP) is violated. In order to avoid a violation of UP the equivalence
principle (EP) is assumed to be violated with the imposition of a
firewall. The unification of QM and GR is still not complete. An elementary
approach to unitarity of black holes prior to the Page time is with
a Bohr-like approach to BH quantum physics {[}8\textendash 10{]},
which will be shortly discussed in next section.

Quantum gravity hair on BHs may be revealed in the collision of two
BHs. This quantum gravity hair on horizons will present itself as
gravitational memory in a GW. This is presented according to the near
horizon condition on Reissnor-Nordstrom BHs, which is $AdS_{2}\times\mathbb{S}^{2}$,
which leads to conformal structures and complementarity principle
between GR and QM.

\section{Bohr-like approach to black hole quantum physics}

At the present time, there is a large agreement, among researchers
in quantum gravity, that BHs should be highly excited states representing
the fundamental bricks of the yet unknown theory of quantum gravitation
{[}8\textendash 10{]}. This is parallel to quantum mechanics of atoms.
In the 1920s the founding fathers of quantum mechanics considered
atoms as being the fundamental bricks of their new theory. The analogy
permits one to argue that BHs could have a discrete energy spectrum
{[}8\textendash 10{]}. In fact, by assuming the BH should be the nucleus
the ``gravitational atom'', then, a quite natural question is: What
are the ``electrons''? In a recent approach, which involves various
papers (see {[}8\textendash 10{]} and references within), this important
question obtained an intriguing answer. The BH quasi-normal modes
(QNMs) (i.e. the horizon's oscillations in a semi-classical approach)
triggered by captures of external particles and by emissions of Hawking
quanta, represent the ``electrons'' of the BH which is seen as being
a gravitational hydrogen atom {[}8\textendash 10{]}. In {[}8\textendash 10{]}
it has been indeed shown that, in the the semi-classical approximation,
which means for large values of the BH principal quantum number $n$,
the evaporating Schwarzschild BH can be considered as the gravitational
analogous of the historical, semi-classical hydrogen atom, introduced
by Niels Bohr in 1913 \cite{key-11,key-12}. Thus, BH QNMs are interpreted
as the BH electron-like states, which can jump from a quantum level
to another one. One can also identify the energy shells of this gravitational
hydrogen atom as the absolute values of the quasi-normal frequencies
{[}8\textendash 10{]}. Within the semi-classical approximation of
this Bohr-like approach, unitarity holds in BH evaporation. This is
because the time evolution of the Bohr-like BH is governed by a time-dependent
Schrodinger equation \cite{key-9,key-10}. In addition, subsequent
emissions of Hawking quanta \cite{key-6} are entangled with the QNMs
(the BH electron states) \cite{key-9,key-10}. Various results of
BH quantum physics are consistent with the results of \cite{key-9,key-10},
starting from the famous result of Bekenstein on the area quantization
\cite{key-13}. Recently, this Bohr-like approach to BH quantum physics
has been also generalized to the Large AdS BHs, \cite{key-14}. For
the sake of simplicity, in this Section we will use Planck units ($G=c=k_{B}=\hbar=\frac{1}{4\pi\epsilon_{0}}=1$).
Assuming that $M$ is the initial BH mass and that $E_{n}$ is the
total energy emitted by the BH when the same BH is excited at the
level $n$ in units of Planck mass (then $M_{p}~=~1$), one gets that
a discrete amount of energy is radiated by the BH in a quantum jump
in terms of energy difference between two quantum levels {[}8\textendash 10{]}
\begin{equation}
\begin{array}{c}
\Delta E_{n_{1}\rightarrow n_{2}}\equiv E_{n_{2}}-E_{n_{1}}=M_{n_{1}}-M_{n_{2}}=\\
=\sqrt{M^{2}-\frac{n_{1}}{2}}-\sqrt{M^{2}-\frac{n_{2}}{2}},
\end{array}\label{eq: jump}
\end{equation}
This equation governs the energy transition between two generic, allowed
levels $n_{1}$ and $n_{2}>n_{1}$ and consists in the emission of
a particle with a frequency $\Delta E_{n_{1}\rightarrow n_{2}}$ {[}8\textendash 10{]}.
The quantity $M_{n}$ in Eq. ($\ref{eq: jump}$), represents the residual
mass of the BH which is now excited at the level $n$. It is exactly
the original BH mass minus the total energy emitted when the BH is
excited at the level $n$ {[}8\textendash 10{]}. Then, $M_{n}=M-E_{n}$,
and one sees that the energy transition between the two generic allowed
levels depends only on the two different values of the BH principal
quantum number and on the initial BH mass {[}8\textendash 10{]}. An
analogous equation works also in the case of an absorption, see {[}8\textendash 10{]}
for details. In the analysis of Bohr \cite{key-11,key-12}, electrons
can only lose and gain energy during quantum jumps among various allowed
energy shells. In each jump, the hydrogen atom can absorb or emit
radiation and the energy difference between the two involved quantum
levels is given by the Planck relation (in standard units) $E=h\nu$.
In the BH case, the BH QNMs can gain or lose energy by quantum jumps
from one allowed energy shell to another by absorbing or emitting
radiation (Hawking quanta). The following intriguing remark finalizes
the analogy between the current BH analysis and Bohr's hydrogen atom.
The interpretation of equation 1 is the energy states of a particle,
that is the electron of the gravitational atom, which is quantized
on a circle of length {[}8\textendash 10{]} 
\begin{equation}
L=4\pi\left(M+\sqrt{M^{2}-\frac{n}{2}}\right).\label{eq: 2}
\end{equation}
Hence, one really finds the analogous of the electron traveling in
circular orbits around the nucleus in Bohr's hydrogen atom. One sees
that it is also  

\begin{equation}
M_{n}=\sqrt{M^{2}-\frac{n}{2}}.\label{eq: 3}
\end{equation}
Thus the uncertainty in a clock measuring a time $t$ becomes, with
the Planck mass is equal to $1$ in Planck units, 
\begin{equation}
\frac{\delta t}{t}=\frac{1}{2M_{n}}=\frac{1}{\sqrt{M^{2}-\frac{n}{2}}},\label{eq: 4}
\end{equation}
which means that the accuracy of the clock required to record physics
at the horizon depends on the BH excited state, which corresponds
to the number of Planck masses it has. More in general, from the Bohr-like
approach to BH quantum physics it emerges that BHs seem to be well
defined quantum mechanical systems, having ordered, discrete quantum
spectra. This issue appears consistent with the unitarity of the underlying
quantum gravity theory and with the idea that information should come
out in BH evaporation, in agreement with a known result of Page \cite{key-7}.
For the sake of completeness and of correctness, we stress that the
topic of this Section, i.e. the Bohr-like treatment of BH quantum
physics, is not new. A similar approach was used by Bekenstein in
1997 \cite{key-15} and by Chandrasekhar in 1998 \cite{key-16}. 

\section{Near Horizon Spacetime and Collision of Black Holes}

This paper proposes how the quantum basis of black holes may be detected
in gravitational radiation. Signatures of quantum modes may exist
in gravitational radiation. Gravitational memory or BMS symmetries
are one way in which quantum hair associated with a black hole may
be detected \cite{key-17}. Conservation of quantum information suggests
that quantum states on the horizon may be emitted or entangled with
gravitational radiation and its quantum numbers and information. In
what follows a toy model is presented where a black hole coalescence
excites quantum hair on the stretched horizon in the events leading
up to the merger of the two horizons. The model is the Poincare disk
for spatial surface in time. To motivate this we look at the near
horizon condition for a near extremal black hole.

The Reissnor-Nordstrom (RN) metric is 
\begin{equation}
ds^{2}~=~-\left(1~-~\frac{2m}{r}~+~\frac{Q^{2}}{r^{2}}\right)dt^{2}~+~\left(1~-~\frac{2m}{r}~+~\frac{Q^{2}}{r^{2}}\right)^{-1}dr^{2}~+~r^{2}d\Omega^{2}.\label{eq: 5}
\end{equation}
Here $Q$ is an electric or Yang-Mills charge and $m$ is the BH mass.
In previous section, considering the Schwarzschild BH, we labeled
the BH mass as $M$ instead. The accelerated observer near the horizon
has a constant radial distance. For the sake of completeness, we recall
that the Bohr-like approach to BH quantum physics has been also partially
developed for the Reissnor-Nordstrom black hole (RNBH) in \cite{key-14}.
In that case, the expression of the energy levels of the RNBH is a
bit more complicated than the expression of the energy levels of the
Schwarzschild BH, being given by (in Planck units and for small values
of $Q$) \cite{key-14} 
\begin{equation}
E_{n}\simeq m-\sqrt{m^{2}+\frac{q^{2}}{2}-Qq-\frac{n}{2}},\label{eq: 6}
\end{equation}
where $q$ is the total charge that has been loss by the BH excited
at the level $n$. Now consider 

\begin{equation}
\rho~=~\int_{r_{+}}^{r}dr\sqrt{g_{rr}}~=~\int_{r^{+}}^{r}\frac{dr}{\sqrt{1~-~2m/r~+~Q^{2}/r^{2}}},\label{eq: 7}
\end{equation}
with lower integration limit $r_{+}$ is some small distance from
the horizon and the upper limit $r$ removed from the black hole.
The result is 
\begin{equation}
\rho~=~m~log[\sqrt{r^{2}~-~2mr~+~Q^{2}}~+~r~-~m]~+~\sqrt{r^{2}~-~2mr~+~Q^{2}}\Big|_{r_{+}}^{r}.\label{eq: 8}
\end{equation}
With a change of variables $\rho~=~\rho(r)$ the metric is

\begin{equation}
ds^{2}~=~\left(\frac{\rho}{m}\right)^{2}dt^{2}~-~\left(\frac{m}{\rho}\right)^{2}d\rho^{2}~-~m^{2}d\Omega^{2},\label{eq: 9}
\end{equation}
where on the horizon $\rho~\rightarrow~r$. This is the metric for
$AdS_{2}\times\mathbb{S}^{2}$ for $AdS_{2}$ in the $(t,~\rho)$
variables tensored with a two-sphere $\mathbb{S}^{2}$ of constant
radius $=~m$ in the angular variables at every point of $AdS_{2}$.
This metric was derived by Carroll, Johnson and Randall \cite{key-18}.

In Section 4 it is shown this hyperbolic dynamics for fields on the
horizon of coalescing BHs is excited. This by the Einstein field equation
will generate gravitational waves, or gravitons in some quantum limit
not completely understood. This GW information produced by BH collisions
will reach the outside world highly red shifted by the tortoise coordinate
$r^{*}~=~r'~-~r~-~2m~ln|1~-~2m/r|$. For a $30$ solar mass BH, which
is mass of some of the BHs which produce gravitational waves detected
by LIGO, the wavelength of this ripple, as measured from the horizon
to $\delta r~\sim~\lambda$ 
\begin{equation}
\delta r'~=~\lambda~-~2m~ln\left(\frac{\lambda}{2m}\right)~\simeq~2\times10^{6}m.
\end{equation}
A ripple in spacetime originating an atomic distance $10^{-10}$m
from the horizon gives a $\nu~=~150Hz$ signal, detectable by LIGO
\cite{key-19}. Similarly, a ripple $10^{-13}$ to $10^{-17}$ $cm$
from the horizon will give a $10^{-1}$Hz signal detectable by the
eLISA interferometer system\cite{key-20}. Thus, quantum hair associated
with QCD and electroweak interactions that produce GWs could be detected.
More exact calculations are obviously required.

Following \cite{key-21}, one can use Hawking's periodicity argument
from the RN metric in order to obtain an ``effective'' RN metric
which takes into account the BH dynamical geometry due to the subsequent
emissions of Hawking quanta. Hawking radiation is generated by a tunneling
of quantum hair to the exterior, or equivalently by the reduction
in the number of quantum modes of the BH. This process should then
be associated with the generation of a gravitational wave. This would
be a more complete dynamical description of the response spacetime
has to Hawking radiation, just as with what follows with the converse
absorption of mass or black hole coalescence. This will be discussed
in a subsequent paper.

These weak gravitons produced by BH hair would manifest themselves
in gravitational memory. The Bondi-Metzner-Sachs (BMS) symmetry of
gravitational radiation results in the displacement of test masses
\cite{key-22}. This displacement requires an interferometer with
free floating mirrors, such as what will be available with the eLISA
system. The BMS symmetry is a record of YM charges or potentials on
the horizon converted into gravitational information. The BMS metric
provide phenomenology for YM gauge fields, entanglements of states
on horizons and gravitational radiation. The physics is correspondence
between YM gauge fields and gravitation. The BHs coalescence is a
process which converts qubits on the BHs horizons into gravitons.

Two BHs close to coalescence define a region between their horizons
with a vacuum similar to that in a Casimir experiment. The two horizons
have quantum hair that forms a type of holographic \textquotedbl{}charge\textquotedbl{}
that performs work on spacetime as the region contracts. The quantum
hair on the stretched horizon is raised into excited states. The ansatz
is made that $AdS_{2}\times\mathbb{S}^{2}$ for two nearly merged
BHs is mapped into a deformed $AdS_{4}$ for a small region of space
between two event horizons of nearly merged BHs. The deformation is
because the conformal hyperbolic disk is mapped into a strip. In one
dimension lower, the spatial region is a two dimensional hyperbolic
strip mapped from a Poincare disk with the same $SL(2,\mathbb{R})$
symmetry. The manifold with genus $g$ for charges has Euler characteristic
$\chi~=~2g~-~2$ and with the $3$ dimensions of $SL(2,~\mathbb{R})$
this is the index $6g~-~6$ for Teichmuller space \cite{key-22}.
The $SL(2,~\mathbb{R})$ is the symmetry of the spatial region with
local charges modeled as a $U(1)$ field theory on an $AdS_{3}$.
The Poincare disk is then transformed into $\mathbb{H}_{p}^{2}$ that
is a strip. The $\mathbb{H}_{p}^{2}~\subset~AdS_{3}$ is simply a
Poincare disk in complex variables then mapped into a strip with two
boundaries that define the region between the two event horizons.

\section{$AdS$ geometry in BH Coalescence}

The near horizon condition for a near extremal black hole approximates
$AdS_{2}\times\mathbb{S}^{2}$. In \cite{key-18} the extremal black
hole replaces the spacelike region in $(r_{+},~r_{-})$ with $AdS_{2}\times\mathbb{S}^{2}$.
For two black holes in near coalescence there are two horizons, that
geodesics terminate on. The region between the horizons is a form
of Kasner spacetime with an anisotropy in dynamics between the radial
direction and on a plane normal to the radial direction. In the appendix
it is shown this is for a short time period approximately an $AdS_{4}$
spacetime. The spatial surface is a three-dimensional Poincare strip,
or a three-dimensional region with hyperbolic arcs. This may be mapped
into a hyperbolic space $H^{3}$. This is a further correlation between
anti-de Sitter spacetimes and black holes, such as seen in $AdS/BH$
correspondences \cite{key-23}.

The region between two event horizons is argued to be approximately
$AdS_{4}$ by first considering the two BHs separated by some distance.
There is an expansion of the area of the $\mathbb{S}^{2}$ that is
then employed with the $AdS_{2}\times\mathbb{S}^{2}$. We then make
some estimates on the near horizon condition for black holes very
close to merging. To start consider the case of two equal mass black
holes in a circular orbit around a central point. We consider the
metric near the center of mass $r~-~0$ and the distance between the
two black holes $d~>>~2m$. In doing this we may get suggestions om
how to model the small region between two black holes about to coalesce.
An approximate metric for two distant black holes is of the form 

\begin{equation}
\begin{array}{c}
ds^{2}~=~\left(1~-~\frac{2m}{|r+d|}~-~\frac{2m}{|r+d|}\right)dt^{2}~\\
\\
-~\left(1~-~\frac{2m}{|r+d|}~-~\frac{2m}{|r+d|}\right)^{-1}dr^{2}~-~r^{2}(d\theta^{2}~+~sin^{2}\theta d\Phi^{2}),
\end{array}
\end{equation}
where $d\Phi~=~d\phi~+~\omega dt$, for $\omega$ the angular velocity
of the two black holes around $r~=~0$. With the approximation for
a moderate Keplerian orbit we may then write this metric as This metric
is approximated with the binomial expansion to $O(r^{2})$ and $O(\omega)$
as 

\begin{equation}
\begin{array}{c}
ds^{2}~=~\left(1~-~\frac{2m}{d}\Big(1~+~2\frac{r^{2}}{d^{2}}\Big)\right)dt^{2}~-~\left(1~-~\frac{2m}{d}\Big(1~+~2\frac{r^{2}}{d^{2}}\Big)\right)^{-1}dr^{2}\\
\\
-~2r^{2}\omega\sin\theta d\phi dt~-~r^{2}(d\theta^{2}~+~sin^{2}\theta d\phi^{2}).
\end{array}
\end{equation}
$g_{tt}$ is similar to the $AdS_{2}$ $g_{tt}$ metric term plus
constant terms and and similarly $g_{rr}$. It is important to note
this approximate metric has expanded the measure of the angular portion
of the metric. This means the $2$-sphere with these angle measures
has more \textquotedblleft area\textquotedblright{} than before from
the contribution of angular momentum. The Ricci curvatures are
\begin{equation}
\begin{array}{c}
R_{tt}~=~R_{rr}~\simeq~-\frac{4m}{d},~R_{t\phi}~\simeq~\left[4\left(1~+~\frac{4m}{d}\right)~+~\frac{16mr^{2}}{d^{3}}\right]\omega sin^{2}\theta,\\
\\
R_{\phi\phi}~=~g_{t\phi}g^{tt}R_{t\phi}~\simeq~-8r^{2}\omega^{2}sin^{4}\theta~+~O\left(\frac{\omega^{2}}{d}\right),~R_{\theta\theta}~=~0,
\end{array}
\end{equation}
where $O(d^{-2})$ terms and higher are dropped. The $R_{r}r$ and
$R_{\phi\phi}$ Ricci curvature are negative and $R_{t\phi}$ positive.
The $2$-surface in $r,~\phi$ coordinates has hyperbolic properties.
This means we have at least the embedding of a deformed version of
$AdS_{3}$ in this spacetime. This exercise expands the boundary of
the disk $\mathbb{D}^{2}$, in a $2$-spacial subsurface, with boundary
around each radial distance so there is an excess angle or \textquotedblleft wedge\textquotedblright{}
that gives hyperbolic geometry. The $(t,~\phi)$ curvature components
comes from the Riemannian curvature $R_{r\phi tr}$ $=~-\frac{1}{2}\omega\alpha^{-1}$
and its contribution to the geodesic deviation equation along the
radial direction is 
\begin{equation}
\frac{d^{2}r}{ds^{2}}~+~R_{\phi tr}^{r}U^{t}U^{\phi}r~=~0
\end{equation}
or that for $U^{t}~\simeq~1$ and $U^{\phi}~\simeq~\omega$ 
\begin{equation}
\frac{d^{2}r}{dt^{2}}~\simeq~\frac{1}{2}\omega^{2}r.
\end{equation}
This has a hyperbolic solution $r~=~r_{0}cosh(\frac{1}{\sqrt{2}}\omega t)$.
The $U^{\phi}$ will have higher order terms that may be computed
in the dynamics for $\phi$ Similarly the geodesic deviation equation
for $\phi$ is 
\begin{equation}
\frac{d^{2}\phi}{ds^{2}}~+~R_{rtr}^{\phi}U^{t}U^{r}r~=~0
\end{equation}
or cryptically 

\begin{equation}
\frac{d^{2}\phi}{dt^{2}}~\simeq~{\bf Riem}~A~cosh(\alpha t)sinh(\alpha t),
\end{equation}
for ${\bf Riem~\rightarrow~R_{rtr}^{\phi}}$. This has an approximately
linear form for small $t$ that turns around into exponential or hyperbolic
forms for larger time. The spatial manifold in the $(r,~\phi)$ variables
then have some hyperbolic structure. It is worth a comment on the
existence of Ricci curvatures for this spacetime. The Schwarzschild
metric has no Ricci curvature as a vacuum solution. This $2$-black
hole solution however is not exactly integrable and so mass-energy
is not localizable. This means there is an effective source of curvature
due to the nonlocalizable nature of mass-energy for this metric. This
argument is made in order to justify the ansatz the spacetime between
two close event horizons prior to coalescence is $AdS_{4}$. Since
most of the analysis of quantum field is in one dimension lower it
is evident there is a subspace $AdS_{3}$. This is however followed
up by looking at geometry just prior to coalescence where the $\mathbb{S}^{2}$
has more area than it can bound in a volume. This leads to hyperbolic
geometry. Above we argue there is an expansion of a disk boundary
$\partial\mathbb{D}^{2}$, and thus hyperbolic geometry. It is then
assume this carries to one additional dimension as well.

Now move to examine two black holes with their horizons very close.
Consider a modification of the $AdS_{2}\times\mathbb{S}^{2}$ metric
with the inclusion of more \textquotedblleft area\textquotedblright{}
in the $\mathbb{S}^{2}$ portion. The addition of area to $\mathbb{S}^{2}$
is then included in the metric. In this fashion the influence of the
second horizon is approximated by a change in the metric of $\mathbb{S}^{2}$.
The metric is then a modified form of the near horizon metric for
a single black hole, 
\begin{equation}
ds^{2}~=~\left(\frac{r}{R}\right)^{2}dt^{2}~-~\left(\frac{R}{r}\right)^{2}dr^{2}~-~(r^{2}~+~\rho^{2})d\Omega^{2}.
\end{equation}
The term $\rho$ means there is additional area to the $\mathbb{S}^{2}$
making it hyperbolic. The Riemann curvatures for this metric are:
\begin{equation}
\begin{array}{c}
R_{trtr}~=~-\frac{1}{r^{2}}~-~\frac{2\rho^{2}}{r^{2}(r^{2}~+~\rho^{2})},~R_{r\theta r\theta}~=~-\frac{\rho^{2}}{r^{2}~+~\rho^{2}},~R_{r\phi r\phi}~=\\
\\
~-\frac{\rho^{2}}{r^{2}~+~\rho^{2}}sin^{2}\theta,~R_{\theta\phi\theta\phi}~=~\rho^{2}sin^{2}\theta
\end{array}
\end{equation}
 From these the Ricci curvatures are 
\begin{equation}
R_{rr}~=~-\frac{1}{r^{2}}~-~2\frac{\rho^{2}}{r^{2}~+~\rho^{2}}^{2},~R_{\theta\theta}~=~R_{\phi\phi}~=~-\left(1~+~\frac{R^{2}}{r^{2}}\right)\frac{\rho^{2}}{r^{2}~+~\rho^{2}},
\end{equation}
that are negative for small values of $r$. For $r~\rightarrow~0$
all Ricci curvatures diverge ${\bf Ric}~\rightarrow~-\infty$. The
$R_{rr}$ diverges more rapidly, which gives this spacetime region
some properties similar to a Kasner metric. However, $R_{rr}~-~R_{\theta\theta}$
is finite for $r~\rightarrow~\infty$. This metric then has properties
of a deformed $AdS_{4}$. With the treatment of quantum fields between
two close horizons before coalescence the hyperbolic space $\mathbb{H}^{2}$
is considered as the spatial surface in a highly deformed $AdS_{3}$.
A Poincare disk is mapped into a hyperbolic strip.

The remaining discussion will now center around the spatial hyperbolic
spatial surface. In particular the spatial dimensions are reduced
by one. This is then a BTZ-like analysis of the near horizon condition.
The $2$ dimensional spatial surface will exhibit hyperbolic dynamics
for particle fields and this is then a model for the near horizon
hair that occurs with the two black holes in this region.

For the sake of simplicity now reduce the dimensions and consider
$AdS_{3}$ in $2$ plus $1$ spacetime. The near horizon condition
for a near extremal black hole in 4 dimensions is considered for the
BTZ black hole. This $AdS_{3}$ spacetime is then a foliations of
hyperbolic spatial surfaces $\mathbb{H}^{2}$ in time. These surfaces
under conformal mapping are a Poincare disk. The motion of a particle
on this disk are arcs that reach the conformal boundary as $t~\rightarrow~\infty$.
This is then the spatial region we consider the dynamics of a quantum
particle. This particle we start out treating as a Dirac particle,
but the spinor field we then largely ignore by taking the square of
the Dirac equation to get a Klein-Gordon wave.

Define the $z$ and $\bar{z}$ of the Poincare disk with the metric
\begin{equation}
ds_{p-disk}^{2}~=~R^{2}g_{z\bar{z}}dzd\bar{z}~=~R^{2}\frac{dzd\bar{z}}{1~-~z\bar{z}}
\end{equation}
with constant negative Gaussian curvature ${\cal R}~=~-4/R^{2}$.
This metric $g_{z\bar{x}}~=~R^{2}/(1~-~\bar{z}z)$ is invariant under
the $SL(2,~\mathbb{R})~\sim$ $SU(1,~1)$ group action, which, for
$g~\in~SU(1,~1)$, takes the form 
\begin{equation}
z~\rightarrow~gz~=~\frac{az~+~b}{\bar{b}z~+~\bar{a}},~g~=~\left(\begin{array}{cc}
a & b\\
\bar{b} & \bar{a}
\end{array}\right).
\end{equation}
The Dirac equation $i\gamma^{\mu}D_{\mu}\psi~+~m\psi~=~0$, $D_{\mu}~=~\partial_{\mu}~+~iA_{\mu}$
on the Poincare disk has the Hamiltonian matrix 
\begin{equation}
{\cal H}~=~\left(\begin{array}{cc}
m & H_{w}\\
H_{w}^{*} & -m
\end{array}\right)
\end{equation}
for the Weyl Hamiltonians 
\begin{equation}
\begin{array}{c}
H_{w}~=~\frac{1}{\sqrt{g_{z\bar{z}}}}\alpha_{z}\left(2D_{z}~+~\frac{1}{2}\partial_{z}(ln~g_{z\bar{z}})\right),\\
\\
H_{w}^{*}~=~\frac{1}{\sqrt{g_{z\bar{z}}}}\alpha_{\bar{z}}\left(2D_{\bar{z}}~+~\frac{1}{2}\partial_{\bar{z}}(ln~g_{z\bar{z}})\right),
\end{array}
\end{equation}
with $D_{z}~=~\partial_{z}~+~iA_{z}$ and $D_{\bar{z}}~=~\partial_{\bar{z}}~+~iA_{\bar{z}}$.
here $\alpha_{z}$ and $\bar{\alpha}_{z}$ are the $2\times2$ Weyl
matrices. 

Now consider gauge fields, in this case magnetic fields, in the disk.
These magnetic fields are topological in the sense of the Dirac monopole
with vanishing Ahranov-Bohm phase. The vector potential for this field
is 
\begin{equation}
A^{\phi}~=~-i\frac{\phi}{2}\left(\frac{dz}{z}~-~\frac{d\bar{z}}{\bar{z}}\right).
\end{equation}
the magnetic field is evaluated as a line integral around the solenoid
opening, which is zero, but the Stokes' rule indicates this field
will be $\phi(\bar{z}-z)/r^{2}$, for $r^{2}~=~\bar{z}z$. A constant
magnetic field dependent upon the volume ${\bf V}~=~\frac{1}{2}dz\wedge d\bar{z}$
in the space with constant Gaussian curvature ${\cal R}~=~-4/R^{2}$
\begin{equation}
{\bf A}^{v}~=~i\frac{BR^{2}}{4}\left(\frac{zd\bar{z}~-~\bar{z}dz}{1~-~\bar{z}z}\right).
\end{equation}
The Weyl Hamiltonians are then 
\begin{equation}
\begin{array}{c}
H_{w}~=~\frac{1~-~r^{2}}{R}e^{-i\theta}\left(\alpha_{z}\Big(\partial_{r}~-~\frac{i}{r}\partial_{\theta}~-~\frac{\sqrt{\ell(\ell~+~1)}~+~\phi}{r}~+~i\frac{kr}{1~-~r^{2}}\Big)\right)\\
\\
H_{w}^{*}~=~\frac{1~-~r^{2}}{R}e^{i\theta}\left(\alpha_{\bar{z}}\Big(\partial_{r}~-~\frac{i}{r}\partial_{\theta}~+~\frac{\sqrt{\ell(\ell~+~1)}~+~\phi}{r}~+~i\frac{kr}{1~-~r^{2}}\Big)\right),
\end{array}
\end{equation}
for $k~=~BR^{2}/4$. With the approximation that $r~<<~1$ or small
orbits the product gives the Klein-Gordon equation 

\begin{equation}
\partial_{t}^{2}\psi~=~R^{-2}\left(\partial_{r}^{2}~+~\frac{\ell(\ell~+~1)~+~\phi^{2}}{r^{2}}~+~k^{2}r^{2}~+~(\ell(\ell~+~1)~+~\phi^{2})k\right)\psi.
\end{equation}
For $\ell(\ell~+~1)~+~\phi^{2}~=~0$ this gives the Weber equation
with parabolic cylinder functions for solutions. The last term $(\ell(\ell~+~1)~+~\phi^{2})k$
can be absorbed into the constant phase 
\begin{equation}
\psi(r,t)~=~\psi(r)e^{-it\sqrt{E^{2}~+~\ell(\ell~+~1)~+~\phi^{2}}}
\end{equation}
This dynamics for a particle in a Poincare disk is used to model the
same dynamics for a particle in a region bounded by the event horizons
of a black hole. With $AdS$ black hole correspondence the field content
of the $AdS$ boundary is the same as the horizon of a black hole.
An elementary way to accomplish this is to map the Poincare disk into
a strip. The boundaries of the strip then play the role of the event
horizons. The fields of interest between the horizons are assumed
to have orbits or dynamics not close to the horizons. The map is $z~=~tanh(\xi)$.
The Klein-Gordon equation is then 
\begin{equation}
\partial_{t}^{2}\psi~=~R^{-2}\left((1~+~2\xi^{2})\partial_{\xi}\partial_{\bar{\xi}}~+~\frac{\ell(\ell~+~1)~+~\phi^{2}}{|\xi|^{2}}~-~k|\xi|^{2}\right)\psi,
\end{equation}
where the $\xi^{2}$ is set to zero under this approximation. The
Klein-Gordon equation is identical to the above. The solution to this
differential equation for $\Phi~=~\ell(\ell~+~1)~+~\phi^{2}$ is 
\begin{equation}
\begin{array}{c}
\psi~=~(2\xi)^{1/4(\sqrt{1~-~4\Phi}~+~1)}e^{-\frac{1}{2}k\xi^{2}}\times\\
\\{}
[c_{1}U\left(\frac{1}{4}\left(\frac{E^{2}R^{2}}{k}~+~\sqrt{1~-~4\Phi}~+~1\right),~\frac{1}{2}(\sqrt{1~-~4\Phi}~+~1),~k\xi^{2}\right)~+\\
\\
~c_{2}L_{\frac{E^{2}R^{2}}{k}+\sqrt{1-4\Phi}}^{\frac{1}{2}\sqrt{1-4\Phi}}(k\xi^{2})].
\end{array}
\end{equation}
The first of these is the confluent hypergeometric function of the
second kind. For $\Phi~=~0$ this reduces to the parabolic cylinder
function. The second term is the associated Laguerre polynomial. The
wave determined by the parabolic cylinder function and the radial
hydrogen-like function have eigenmodes of the form in the diagram
above. The parabolic cylinder function $D_{n}~=~2^{n/2}e^{-x^{2}/4}H_{n}(x/\sqrt{2})$
with integer $n$ gives the Hermite polynomial. The recursion formula
then gives the modes for the quantum harmonic oscillator. The generalized
Laguerre polynomial $L_{n-\ell-1}^{2\ell+1}(r)$ of degree $n~-~\ell~-~1$
gives the radial solutions to the hydrogen atom. The associated Laguerre
polynomial with general non-integer indices has degree associated
with angular momentum and the magnetic fields. This means a part of
this function is similar to the quantum harmonic oscillator and the
hydrogen atom. The two parts in a general solution have amplitudes
$c_{1}$ and $c_{2}$ and quantum states in between the close horizons
of coalescing black holes are then in some superposition of these
types of quantum states, see Figure 1. 

\includegraphics[scale=0.75]{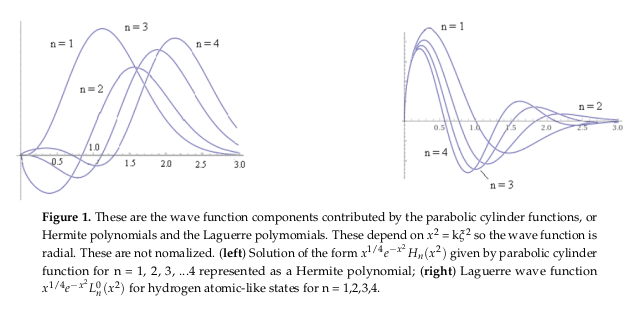}

The Hamiltonian 

\begin{equation}
H~=~\frac{1}{2}|\pi|^{2}~-~\frac{g}{r^{2}},~\pi~=~-i\partial_{r},
\end{equation}
which contains the monopole field, describes the motion of a gauge
particle in the hyperbolic space. In addition, there is a contribution
from the constant magnetic field $U~=~-kr^{2}/2$. Now convert this
theory to a scalar field theory with $r~\rightarrow~\phi$ and $\pi~=~-i\partial_{r}\phi$.
Finally introduce the dilaton operator $D$ and the scalar theory
consists of the operators 
\begin{equation}
H_{0}~=~\frac{1}{2}|\pi|^{2}~-~\frac{g}{\phi^{2}},~U~=~-k\frac{\phi^{2}}{2},~D~=~\frac{1}{4}(\phi\pi~+~\pi\phi),
\end{equation}
where $H_{0}~+~U$ is the field theoretic form of the potential in
equation 9. These potentials then lead to the algebra 

\begin{equation}
[H_{0},~U]~=~-2iD,~[H_{0},~D]~=~-iH_{0},~[U,~D]~=~iM.
\end{equation}
This may be written in a more compact form with $L_{0}~=~2^{-1/2}(H_{0}~+~U)$,
which is the total Hamiltonian, and $L_{\pm}~=~2^{-1/2}(U~-~H_{0}~\pm~iD)$.
This leaves the $SL(2,~\mathbb{R})$ algebra 
\begin{equation}
[L_{0},~L_{\pm}]~=~\pm iL_{0},~[L_{+},~L_{-}]~=~L_{0}.
\end{equation}
This is the standard algebra $\sim~{\mathfrak{s}\mathfrak{u}}(2)$.
Given the presence of the dilaton operator this indicates conformal
structure. The space and time scale as $(t,~x)~\rightarrow~\lambda(t,~x)$
and the field transforms as $\phi~\rightarrow~\lambda^{\Delta}\phi$.
The measure of the integral $d^{4}x\sqrt{g}$ is invariant, where
$\lambda~=~\partial x'/\partial x$ gives the Jacobian $J~=~det|\frac{\partial x'}{\partial x}|$
that cancels the $\sqrt{g}$ and the measure is independent of scale.
In doing this, we are anticipating this theory in four dimensions.
We then simply have the scaling $\phi~\rightarrow~\lambda^{-1}\phi$
and $\pi~\rightarrow~\pi$. For the potential term $-g/2\phi^{2}$
invariance of the action requires $g~\rightarrow~\lambda^{-2}g$ and
for $U~=~-k\frac{\phi^{2}}{2}$ clearly $k~\rightarrow~\lambda^{2}k$.
This means we can consider this theory for $2$ space plus $1$ time
and its gauge-like group $SL(2,~\mathbb{R})$ as one part of an $SL(2,~\mathbb{C})~\sim~SL(2,~\mathbb{R})^{2}$.

The differential equation number 28 is a modified form of the Weber
equation $\psi_{xx}~-~(\frac{1}{4}x^{2}~+~c)\psi~=~=0.$ The solution
in Abramowit and Stegun are parabolic cylinder functions $D_{-a-1/2}(x)$,
written according to hypergeometric functions. The $\xi^{-1}$ part
of the differential equation contributes to the Laguerre polynomial
solution. If we let $\xi~=~e^{x/2}$ and expand to quadratic powers
we then have the potential in the variable x

\begin{equation}
V(x)~=~-(g~+~k)~+~\frac{1}{2}(k~-~g)(x^{2}~+~x^{4}),
\end{equation}
for $g$ and $k$ the constants in $H_{0}$ and $U$. The Schrodinger
equation for this potential with a stationary phase in time has the
parabolic cylinder function solution

\[
\psi(x)~=~c_{1}D_{\frac{\beta^{2}-4(\alpha+2\sqrt{2}\alpha^{3/2})}{16\sqrt{2}\alpha^{3/2}}}\left(\frac{\beta(1+4x)}{\sqrt{2}(2\alpha)^{3/4}}\right)~+~c_{2}D_{\frac{-\beta^{2}-4(\alpha-2\sqrt{2}\alpha^{3/2})}{16\sqrt{2}\alpha^{3/2}}}\left(\frac{i\beta(1+4x)}{\sqrt{2}(2\alpha)^{3/4}}\right),
\]
where $\alpha~=~g~+~k$ and $\beta~=~k~-~g$. The parabolic cylinder
function describes a theory with criticality, which in this case has
with a Ginsburg-Landau potential. The field theory form also has parabolic
cylinder function solutions. The field theory with the field expanded
as $\phi~=~e^{\chi}$ is expanded around unity so $\phi~\simeq$ $1~+~\chi~+~\frac{1}{2}\chi^{2}$.
A constant $C$ such that $\chi~\rightarrow~C\chi$ is unitless is
assumed or implied to exist. The Lagrangian for this theory is

\begin{equation}
{\cal L}~=~\frac{1}{2}\partial_{\mu}\chi\partial^{\mu}\chi~+~\alpha~+~\frac{1}{2}\mu^{2}\chi^{2}~+~2\beta\mu\chi.
\end{equation}
The constant $\mu$, standing for mass and absorbing $\alpha$, is
written for dimensional purposes. We then consider the path integral
$Z~=~D[\chi]e^{-iS-i\chi J}$. Consider the functional differentials
acting on the path integral

\begin{equation}
\left((p^{2}~+~m^{2})\frac{\delta}{\delta J}~-~2i\beta\right)Z~=~-i\left\langle \frac{\delta S}{\delta\chi}\right\rangle ,
\end{equation}
where $\partial_{\mu}\chi~=~p_{\mu}\chi$. The Dyson-Schwinger theorem
tells us that $\left\langle \frac{\delta S}{\delta\chi}\right\rangle ~=~\langle J\rangle$
mean we have a polynomial expression $\langle\frac{1}{2}(p^{2}~+~m^{2})\chi$
$-~i\beta~-~J\rangle~=~0$, where we can trivially let $J~-~i\beta~\rightarrow~J$.
This does not lead to parabolic cylinder functions. There has been
a disconnect between the ordinary quantum mechanical theory and the
QFT. We may however, continue the expansion to quartic terms. This
will also mean there is a cubic term, we may impose that only the
real functional variation terms contribute and so only even power
of the field define the Lagrangian 
\begin{equation}
{\cal L}~\rightarrow~\frac{1}{2}\partial_{\mu}\chi\partial^{\mu}\chi~+~\alpha~+~\frac{1}{2}\mu^{2}\chi^{2}~+~\frac{1}{4}\lambda\chi^{4},
\end{equation}
where $\frac{2}{3}\alpha~\rightarrow~\frac{1}{4}\lambda$. The functional
derivatives are then 
\begin{equation}
\left((p^{2}~+~m^{2})\frac{\delta}{\delta J}~+~\lambda\frac{\delta^{3}}{\delta J^{3}}\right)Z~=~-i\left\langle \frac{\delta S}{\delta\chi}\right\rangle ,
\end{equation}
This cubic form has three parabolic cylinder solutions. We may think
of this as $ap~+~bp^{3}~=~J$ and is a cubic equation for the source
$J$ that is annulled at three points. The correspond to distinct
solutions with distinct paths. These three solutions correspond to
three contours and define three distinct vacua. The overall action
is a quartic function, which will have three distinct vacua, where
one of these is the low energy physical vacua. It is worth noting
this transformation of the problem has converted it into a system
similar to the Higgs field. This system with both harmonic oscillator
and a Coulomb potentials is conformal and it maps into a system with
parabolic cylinder functions solutions. In effect there is a transformation
$harmonic~oscillator$ $states$ $\leftrightarrow$ $hydrogen-like~states$.
The three solutions would correspond to the continuance of conformal
symmetry, but where the low energy vacuum for one of these may not
appear to be conformally invariant.

This scale transformation above is easily seen to be the conformal
transformation with $\lambda~=~\Omega$. The scalar tensor theory
of gravity for coupling constant $\kappa~=~16\pi G$ 
\begin{equation}
\begin{array}{c}
S[g,~\phi]~=~\int d^{4}x\sqrt{g}\left(\frac{1}{\kappa}R~+~\frac{1}{2}\partial_{\mu}\phi\partial^{\mu}\phi~+~V(\phi)\right),\\
\\
S[g,~\phi]~=~\int d^{4}x\sqrt{g}\left(\frac{1}{\kappa}R~+~\frac{1}{2}\partial_{\mu}\phi\partial^{\mu}\phi~+~V(\phi)\right).
\end{array}
\end{equation}
This then has the conformal transformations 

\begin{equation}
g'_{\mu\nu}~=~\Omega^{2}g_{\mu\nu},~\phi'~=~\Omega^{-1}\phi,~\Omega^{2}~=~1~+~\kappa\phi^{2},
\end{equation}
with the transformed action 
\begin{equation}
S[g',~\phi']~=~\int d^{4}x\sqrt{g'}\left(\frac{1}{\kappa}R'~+~\frac{1}{2}g'^{\mu\nu}\partial_{\mu}\phi'\partial_{\nu}\phi'~+~V(\phi')~+~\frac{1}{12}R\phi'^{2}\right).
\end{equation}
There is then a hidden $SO(3,~1)~\simeq~SL(2,~\mathbb{C})$ symmetry.
Given an internal index on the scalar field $\phi^{i}$ there is a
linear $SO(n)$ transformation $\delta\phi^{i}~=$ $C^{ijk}\phi_{j}\delta\tau_{k}$
for $\tau_{k}$ a parameter. There is also a nonlinear transformation
from equation 12 as $\delta\phi^{i}~=~(1~+~\kappa\phi^{2})^{1/2}\kappa\delta\chi^{i}$
for $\chi^{i}$ a parameterization. In the primed coordinates the
scalar field and metric transform as

\begin{equation}
\begin{array}{c}
\delta\phi^{i}~=~\delta\tau^{i}~-~\kappa\phi'^{i}\phi^{j}\delta\chi^{j}\\
\\
\delta g_{\mu\nu}~=~\frac{2g'_{\mu\nu}\kappa\phi'^{i}\delta\chi^{i}}{1~-~\kappa\phi'^{2}}.
\end{array}
\end{equation}
The gauge-like dynamics have been buried into the scalar field. With
this semi-classical model the scalar field adds some renormalizability.
Further this model is conformal. The conformal transformation mixes
the scalar field, which is by itself renormalizable, with the spacetime
metric. Quantum gravitation is however difficult to renormalize. Yet
we see the linear group theoretic transformation of the scalar field
in $SO(n)$ is nonlinear in $SO(n,~1)$.

Conformal symmetry is manifested in sourceless spacetime, or spatial
regions without matter or fields. The two dimensional spatial surface
in $AdS_{3}$ is the Poincare disk that with complexified coordinates
has metric with $SL(2,~\mathbb{R})$ algebraic structure. This may
of course be easily extended into $SL(2,~\mathbb{C})$ as $SL(2,~\mathbb{R})\times SL(2,~\mathbb{R})$.
In this conformal setting quantum states share features similar to
the emission of photons by a harmonic oscillator or an atom. The orbits
of these paths are contained in regions bounded by hyperbolic surfaces,
or arcs for the two dimensional Poincare disk. The entropy associated
with these arcs is a measure of the area contained within these curves.
This is in a nutshell the Mirzakhani result on entropy for hyperbolic
curves.

This development is meant to illustrate how radiation from black holes
is produced by quantum mechanical means not that different from bosons
produced by a harmonic oscillator or atom. Hawking radiation in principle
is detected with a wavelength not different from the size of the black
hole. The wavelength approximately equal to the Schwarzschild radius
has energy $E~=~h\nu$ corresponding to a unit mass emitted. The mass
of the black hole is $n$ of these units and it is easy to find $m_{p}~=$
$\sqrt{\hbar c/G}$. These modes emitted are Planck units of mass-energy
that reach ${\cal I}^{\infty}$. In the case of gravitons, these carry
gravitational memory. For the coalescence of black holes gravitational
waves are ultimately gravitons. For Hawking radiation there is the
metric back reaction, which in a quantum mechanical setting is an
adjustment of the black hole with the emission of gravitons. The emission
of Hawking radiation might then be compared to a black hole quantum
emitting a Planck unit of black hole that then decays into bosons.
The quantum induced change in the metric is a mechanism for producing
gravitons.

In the coalescence of black holes the quantum hair on the stretched
horizons sets up a type of Casimir effect with the vacuum that generates
quanta. In general these are gravitons. We might see this as not that
different from a scattering experiment with two Planck mass black
holes. These will coalesce, form a larger black hole, produce gravitons,
and then quantum states excited by this process will decay. The production
of gravitons by this mechanism is affiliated with normal modes in
the production of gravitons, which in principle is not different from
the production of photons and other particles by other quantum mechanical
processes. I fact quantum mechanical processes underlying black hole
coalescence might well be compared to nuclear fusion.

The 2 LIGOs plus now the VIRGO detector are recording and triangulating
the positions of distant black hole collisions almost weekly. This
information may contain quantum mechanical information associated
with quantum gravitation. This information is argued below to contain
BMS symmetries or information. This will be most easily detected with
a space based system such as eLISA, where the shift in metric positions
of test masses is most readily detectable. However, preliminary data
with the gross displacement of the LIGO mass may give preliminary
information as well.

\section{Discussion}

The coalescence of two black holes is a form of scattering. We may
think of black holes as an excited state of the quantum gravity field
and a sort of elementary particle. The scattering of two black holes
results in a larger black hole plus gravitational radiation. This
black hole will then emit Hawking radiation. Thus in general the formation
of black holes, their coalescence and ultimate quantum evaporation
can be seen as intermediate processes in a general scattering theory. 

Quantum hair is a set of quantum fields that build up quantum gravitation,
in the manner of gauge-gravity duality and BMS symmetry. This is holography,
with the fields on the horizons of two BHs that determine the graviton/GW
content of the BH coalescence. A detailed analysis of this may reveal
BMS charges that reach ${\cal I}^{+}$ are entangled with Hawking
radiation by a form of entanglement swap. In this way Hawking radiation
may not be entangled with the black hole and thus not with previously
emitted Hawking radiation. This will be addressed later, but a preliminary
to this idea is seen in \cite{key-24}, for disentanglement between
Hawking radiation and a black hole. The authors are working on current
calculations where this is an entanglement swap with gravitons. The
black hole production of gravitons in general is then a manifestation
of quantum hair entanglement.

It is illustrative for physical understanding to consider a linearized
form of gravitational memory. Gravitational memory from a physical
perspective is the change in the spatial metric of a surface according
to \cite{key-4} 
\begin{equation}
\Delta h_{+.\times}~=~\lim_{t\rightarrow\infty}h_{+,\times}(t)~-~\lim_{t\rightarrow-\infty}h_{+,\times}(t).
\end{equation}
Here $+$ and $\times$ refer to the two polarization directions of
the GW. See \textbackslash{}cite\{key-20\} for more on this. Quantum
hair on two black holes just before coalescence are highly excited
and contribute to spacetime curvature, or in a full context of quantum
gravitation the generation of gravitons. As yet there is no complete
theory of quantum gravity, but it is reasonable to think of gravitational
radiation as a classical wave built from many gravitons. Gravitons
have two polarizations and a state $|\Psi_{+,\times}\rangle$ the
density matrix $\rho_{+,\times}~=$ $|\Psi_{+,\times}\rangle\langle\Psi_{+,\times}|$
then defines entropy $S~=~\rho_{+,\times}log(\rho_{+,\times})$ that
with this near horizon condition of $AdS$ with a black hole is a
form of Mirzakhani entropy measure in hyperbolic space. The gravitons
emitted are generated by quantum hair on the colliding black holes.
These will contribute to gravitational waves, and in general with
BMS translations that bear quantum information from quantum hair.

This theory connects to fundamental research, The entanglement entropy
of $CFT_{2}$ entropy with $AdS_{3}$ lattice spacing $a$ is 
\begin{equation}
S~\simeq~\frac{R}{4G}ln(|\gamma|)~=~\frac{R}{4G}ln\left[\frac{\ell}{L}~+~e^{2\rho_{c}}sin\left(\frac{\pi\ell}{L}\right)\right],
\end{equation}
where the small lattice cut off avoids the singular condition for
$\ell~=~0$ or $L$ for $\rho_{c}~=~0$. For the metric in the form
$ds^{2}~=~(R/r)^{2}(-dt^{2}~+~dr^{2}~+~dz^{2})$ the geodesic line
determines the entropy as the Ryu-Takayanagi (RT) result \cite{key-1}
\begin{equation}
\begin{array}{c}
S~=~\frac{R}{2G}\int_{2\ell/L}^{\pi/2}\frac{ds}{sin~s}~=~-\frac{R}{2G}ln[cot(s)~+~csc(s)]\Big|_{2\ell/L}^{\pi/2}\\
\\
\simeq~\frac{R}{2G}ln\left(\frac{\ell}{L}\right),
\end{array}
\end{equation}
which is the small $\ell$ limit of the above entropy. The RT result
specifies entropy, which is connected to action $S_{a}~\leftrightarrow~S_{e}$
\cite{key-25}. Complexity, a form of Kolmogoroff entropy \cite{key-26},
is $S_{a}/\pi\hbar$ which can also assume the form of the entropy
of a system $S~\sim~k~log(dim~\{{\cal H}\})$ for ${\cal H}$ the
Hilbert space and the dimension over the number of states occupied
in the Hilbert space. There is also complexity as the volume of the
Einstein-Rosen bridge \cite{key-27} $vol/GR_{ads}$ or equivalently
the RT area $\sim~vol/R_{AdS}$. There is an equivalency between entropy
or complexity according to the geodesic paths in hyperbolic $\mathbb{H}^{2}$
by geometric means \cite{key-22}. This should generalize to $\mathbb{H}^{3}$
$\subset~AdS_{4}$.

The generation of gravitational waves should have an underlying quantum
mechanical basis. It is sometimes argued that spacetime physics may
not be at all quantum mechanical. This is probably a good approximation
for energy sufficient orders of magnitude lower than the Planck scale.
However, if we have a scalar field that define the metric $g'~=~g'(g,~\phi)$
with action $S[g,~\phi]$ then a quantum field $\phi$ and a purely
classical $g$ means the transformation of $g$ by this field has
no quantum physics. In particular for a conformal theory $\Omega~=~1~+~\kappa\phi^{a}\phi^{a}$,
here $a$ an internal index, the conformal transformation $g'_{\mu\nu}~=~\Omega^{2}g_{\mu\nu}$
has no quantum content. This is an apparent inconsistency. For the
inflationary universe the line element 

\begin{equation}
ds'^{2}~=~g'_{\mu\nu}dx^{\mu}dx^{\nu}~=~\Omega^{2}(du^{2}~-~d\Sigma^{(3)})
\end{equation}
with $dt/du~=~\Omega^{2}$ gives an FLRW or de Sitter-like line element
that expands space with $\Omega^{2}~=~e^{t\sqrt{\Lambda/3}}$. The
current slow accelerated universe we observe is approximately of this
nature. The inflaton scalars are then fields that stretch space as
a time dependent conformal transformation and are quantum mechanical.

The generation of gravitational waves is ultimately the generation
of gravitons. Signatures of these quantum effects in black hole coalescence
will entail the measurement of quantum information. Gravitons carry
BMS charges and these may be detected with a gravitational wave interferometer
capable of measuring the net displacement of a test mass. The black
hole hair on the stretched horizon is excited by the merger and these
results in the generation of gravitons. The Weyl Hamiltonians in equation
9 depend on the curvature as $~\propto~\sqrt{{\cal R}}$. For the
curvature extreme during the merging of black holes this means many
modes are excited. The two black holes are pumped with energy by the
collision, this generates or excites more modes on the horizons, where
this results in a black hole with a net larger horizon area. This
results in a metric response, or equivalently the generation of gravitons.

Quantum normal modes are given by independent eigen-states, such as
with quantum harmonic oscillator states. The harmonic oscillator states
are well known to be given by the Hermite polynomials, which are a
special case of parabolic cylinder functions. Rydberg states are also
a form of normal modes. The quantum states for the hyperbolic geometry
of black hole mergers are a generalization of these forms of states.
The excitation of quantum hair in such a merger and the production
of gravitons is a converse situation for the emission of Hawking radiation.
In both cases there is a dynamical response of the metric, which is
associated with gravitons. Currently a \textquotedblleft by hand\textquotedblright{}
correction called back reaction is used in models. A more explicit
discussion on the production of gravitons is beyond the scope here.
However, the parabolic cylinder functions and the Laguerre functions
clearly play a role in quantum production of gravitons in BH coalescence.
This means quantum gravitation should have signatures of much the
same physics as atomic physics or the role of electrons and phonons
in solids.

The major import of this expository is to propose quantum gravitational
signatures in the coalescence of black holes. While there is plenty
of further development needed to compute more firm predictions, the
generic result is that gravitational waves from colliding black holes
have some quantum gravitational signatures. These signatures are to
be found in gravitational memory. Further, this long-term adjustment
of spacetime metric deviates form a purely classical expected result.
With further advances in gravitational wave interferometry, in particular
with the future eLISA space mission, it should be possible to detect
elements of gravitons and quantum gravitation.

\end{document}